\documentclass[12pt]{article}
\usepackage{graphicx}

\newcommand\pubnumber{DPF2015-220}
\newcommand\UCPreprint{UCHEP-15-05}
\newcommand\pubdate{\today}

\def\cincinnati{Department of Physics, University of Cincinnati,
  Cincinnati, Ohio 45221, US}

\def\Title#1{\begin{center} {\Large #1 } \end{center}}
\def\Author#1{\begin{center}{ \sc #1} \end{center}}
\def\Address#1{\begin{center}{ \it #1} \end{center}}

\newcommand\pubblock{\rightline{\begin{tabular}{l} \pubnumber\\
      \UCPreprint \\
         \pubdate  \end{tabular}}}
\newenvironment{Abstract}{\begin{quotation}  }{\end{quotation}}
\newenvironment{Presented}{\begin{quotation} \begin{center} 
             PRESENTED AT\end{center}\bigskip 
      \begin{center}\begin{large}}{\end{large}\end{center} \end{quotation}}
\def\Acknowledgments{\bigskip  \bigskip \begin{center} \begin{large}
             \bf ACKNOWLEDGMENTS \end{large}\end{center}}

\def\beq{\begin{equation}}
\def\eeq#1{\label{#1}\end{equation}}
\def\eeqn{\end{equation}}

\def\beqa{\begin{eqnarray}}
\def\eeqa#1{\label{#1}\end{eqnarray}}
\def\eeqan{\end{eqnarray}}

\let\bar=\overbar

\def\Dslash{\not{\hbox{\kern-4pt $D$}}}
\def\dslash{\not{\hbox{\kern-2pt $\del$}}}

\def\msb{{\bar{\ssstyle M \kern -1pt S}}}

\begin{document}
\begin{titlepage}
\pubblock

\vfill
\Title{Searches for New Physics at the Belle II Experiment}
\vfill
\Author{Wang, Boqun}
\Address{\cincinnati}
\vfill
\begin{Abstract}
  The Belle II experiment at the SuperKEKB collider is an upgrade of
  the Belle / KEKB experiment. It will start physics data taking from
  2018 and with $\sim 40$ times luminosity, its goal is to accumulate
  50 $ab^{-1}$ of $e^+e^-$ collision data. The physics programs
  have a wide range of areas for new physics, such as more constraints
  on CKM Unitarity Triangle, searching for charged Higgs, direct CPV,
  Lepton Flavour Violation and dark matter.In this monograph, we will
  review the current status of Belle II and SuperKEKB construction and
  introduce the main physics opportunities at this facility.
\end{Abstract}
\vfill
\begin{Presented}
DPF 2015\\
The Meeting of the American Physical Society\\
Division of Particles and Fields\\
Ann Arbor, Michigan, August 4--8, 2015\\
\end{Presented}
\vfill
\end{titlepage}
\def\thefootnote{\fnsymbol{footnote}}
\setcounter{footnote}{0}

\section{Introduction}

B factory is an $e^+e^-$ collider which runs at the $\Upsilon(4S)$
resonance energy to produce B meson pairs. The major B factories are
Belle running at KEKB and BaBar running at PEP-II. They have totally
collected $~ 1.5 ab^{-1}$ of $e^+e^-$ collision data. With that data
sample, they've reached fruitful physics achievements in a wide range
of areas, such as the CKM angle measurement, $|V_{cb}|$ and $V_{ub}$
measurement, semileptonic and leptonic B decays, rare B decays, $\tau$
physics, $D^0$ mixing and CPV, B physics at the $\Upsilon(5S)$,
two-photon physics and new resonances.

Belle II experiment at SuperKEKB collider is an upgrade of Belle to search
for New Physics, which is physics beyond the Standard Model (SM), by
plan to take $~ 50 ab^{-1}$ $e^+e^-$ collision data. The SuperKEKB
asymmetric electron positron collider can provide a clean environment
for producing B meson pairs via $\Upsilon(4S)$ resonance decay. Its
designed luminosity is $8 \times 10^{35} cm^{-2} s^{-1}$, which is
about 40 times larger than the KEKB collider. The 50 $ab^{-1}$ overall
integrated luminosity corresponds to 55 billion of $B \overline{B}$
pairs, 47 billion of $\tau^+\tau^-$ pairs and 65 billion $c
\overline{c}$ states.

In this article, we will introduce the Belle II / SuperKEKB
experiment, the opportunities for new physics on it, and the current
status and future plan of the experiment.

\section{Belle II / SuperKEKB}

For achieving the 40 times luminosity compared with KEKB, many parts
of the accelerator are upgraded. The most important part is the beam
size. The beam bunches are significant squeezed to obtain the
so-called nano-beam. The beam energies will be changed slightly to
have a less boosted center-of-mass system.

\begin{figure}[htb]
\centering
\includegraphics[width=0.8\textwidth]{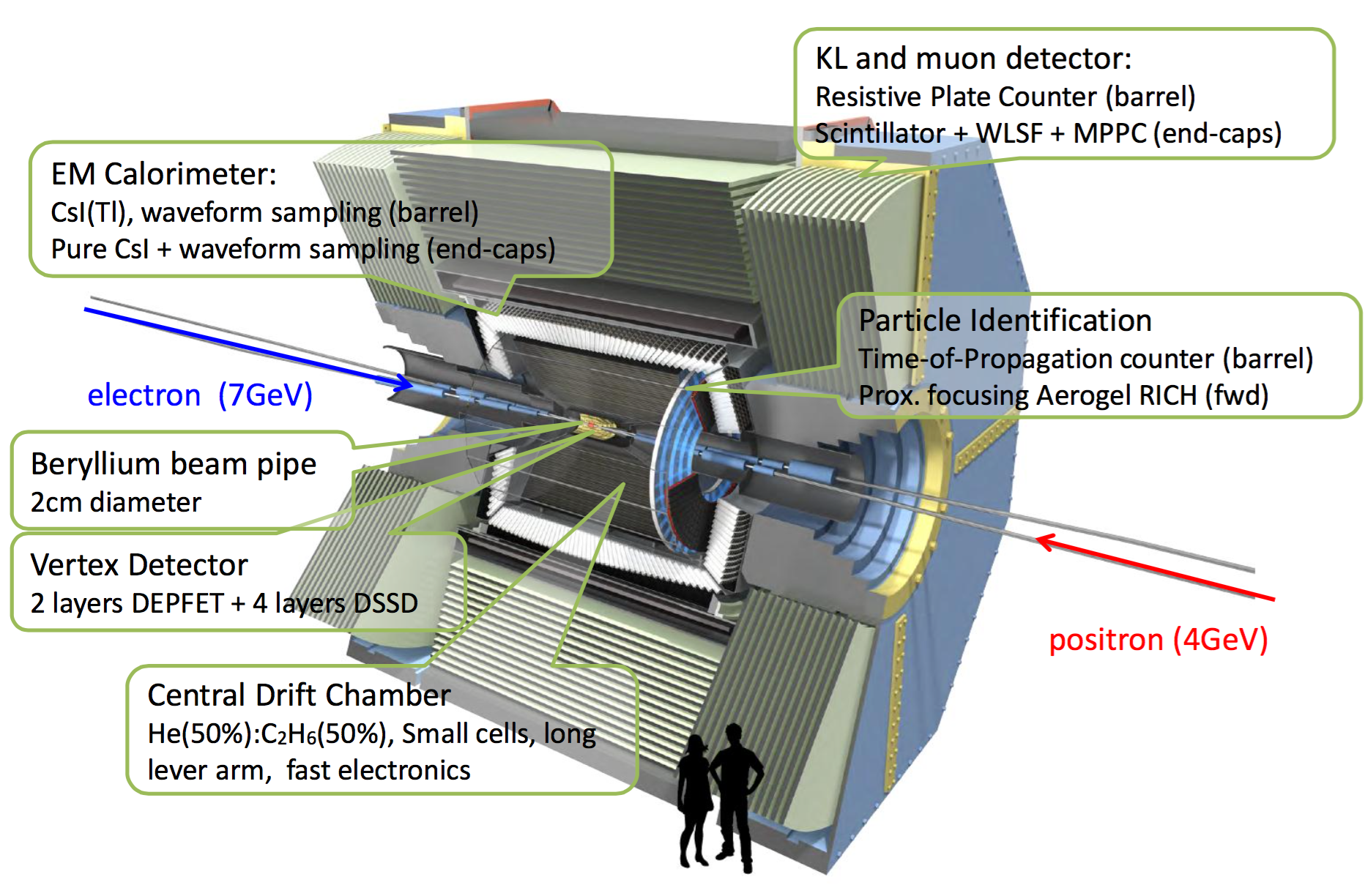}
\caption{The Belle II detector.}
\label{fig:belle-ii}
\end{figure}

Most subdetectors of Belle will be upgraded accordingly for Belle II
apparatus, as shown in Figure~\ref{fig:belle-ii}. This includes the
newly designed vertex detection system (PXD and SVD), a drift chamber
with longer arms and smaller cells, a completely new PID system which
consists of TOP detector at the barrel and ARICH detector in the
forward end, the electro-magnetic calorimeter (ECL) with upgraded
crystals and electronics, and upgraded $K_L$-$\mu$ detection system
(KLM). The trigger system, DAQ, and software systems are also upgraded
accordingly. With the high radiation intensity comes with the high
luminosity, the Belle II detector has to be capable of handling higher
beam related backgrounds.

\section{Opportunities for New Physics on Belle II}

With much larger data set, there are lots of opportunities for New
Physics searches on Belle II.

\subsection{Constraining the CKM UT}

\begin{figure}[htb]
\centering
\includegraphics[width=0.8\textwidth]{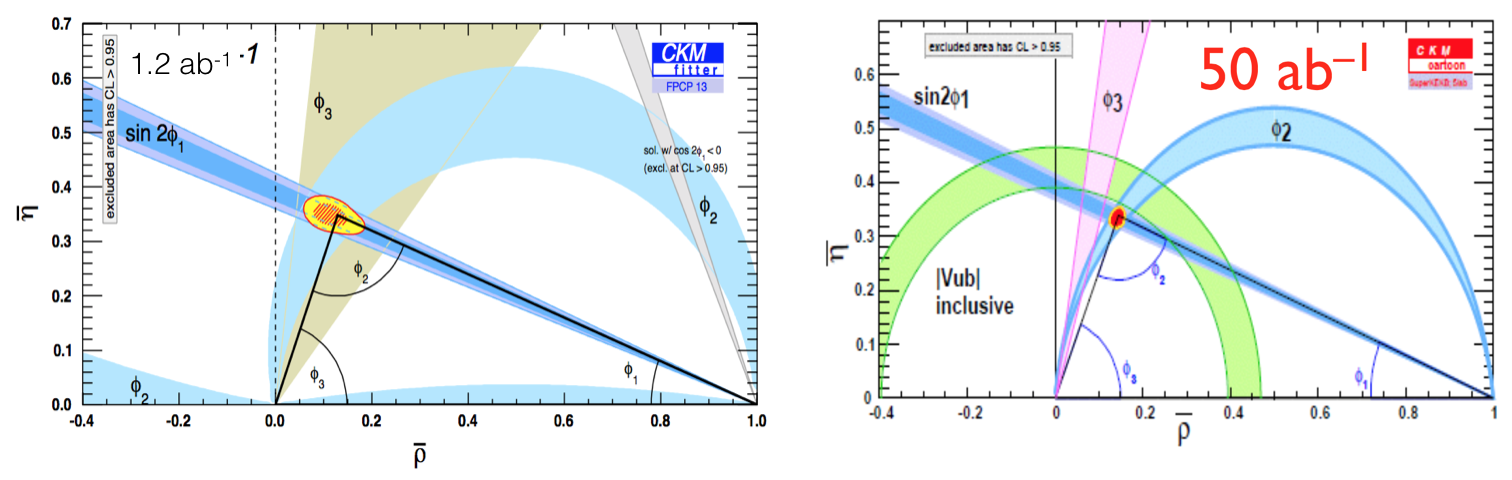}
\caption{Comparison of the current precision of CKM UT (left) and the
  predicted precision with the 50 $ab^{-1}$ data collected with Belle
  II (right).}
\label{fig:ckm-comp}
\end{figure}

With higher integrated luminosity, the precision of CKM Unitarity
Triangle (UT) parameters could be significantly improved, as shown in
Figure~\ref{fig:ckm-comp}.

\subsection{$b \rightarrow s \gamma$ Decays}

\begin{figure}[htb]
\centering
\includegraphics[width=0.8\textwidth]{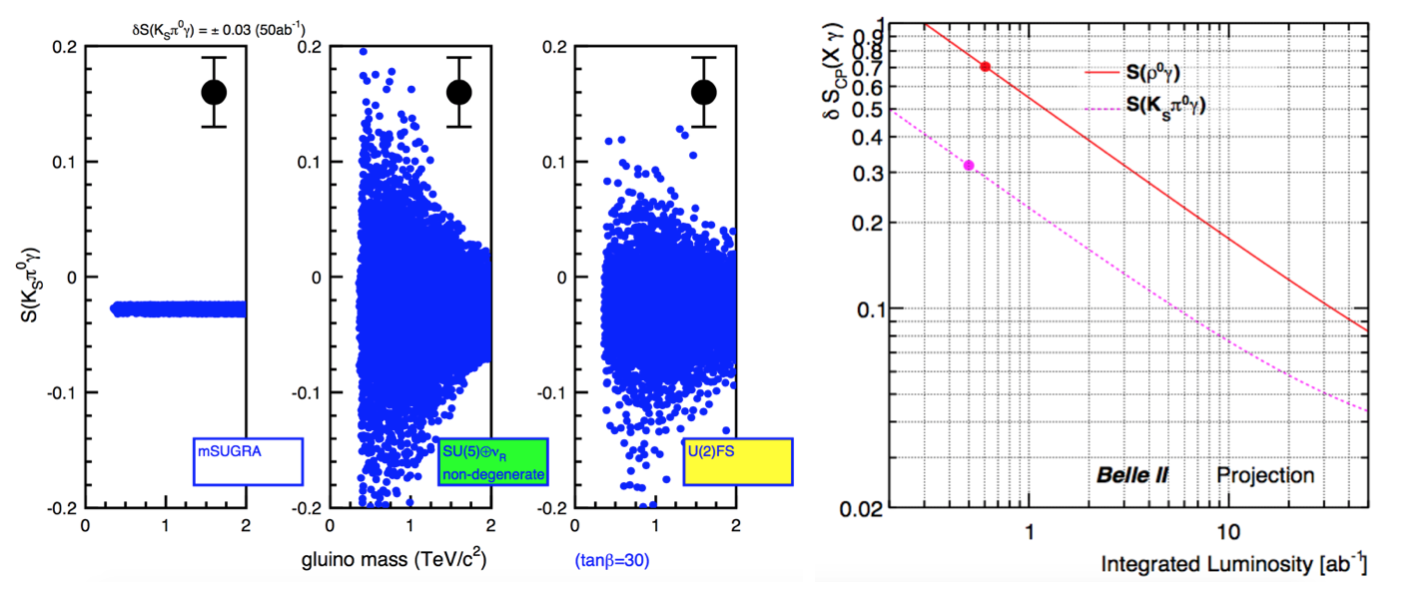}
\caption{The asymmetry predicted by different SUSY models (left) and
  the measurement precision with Belle II data set (right).}
\label{fig:bsg}
\end{figure}

In Standard Model (SM), the decay $b \rightarrow s \gamma$ is
suppressed. The CP asymmetry for $B^0 \rightarrow K_S \pi^0 \gamma$ is
predicted as about 0.04. With the precision predicted with the large
data set from Belle II, it is possible to distinguish between
different SUSY models by measuring the asymmetry~\cite{bsg}, as shown in
Figure~\ref{fig:bsg}. 

\subsection{Charged Higgs: $B^+ \rightarrow \tau^+ \nu$ and $B \rightarrow D^{(*)} \tau \nu$}

The decays $B^+ \rightarrow \tau^+ \nu$~\cite{btaunu} and $B \rightarrow D^{(*)}
\tau \nu$ are very sensitive to a charged Higgs. In some NP models,
charged Higgs can be exchanged in addition to the W boson. The
branching fractions of these decays could be enhanced, and this can be
checked by using the large data sample from Belle II. 

\subsection{Direct CPV: $B \rightarrow K \pi$}

Standard Model predicts the CP asymmetry difference $A_{CP}$ between
$B^0 \rightarrow K^+\pi^-$ and $B^+ \rightarrow K^+\pi^0$ should be
zero, but the measurement of the $K\pi$ channels shows that $A_{CP}$
is not zero~\cite{bkpi}.


The independent sum rule predicts that~\cite{bkpi-sumrule}:

\[
A_{CP}^{K^+\pi^-} + A_{CP}^{K^0\pi^+}\frac{B(B^+ \rightarrow
  K^0\pi^+)\tau_{B^0}}{B(B^0 \rightarrow K^+\pi^-)\tau_{B^+}} =
A_{CP}^{K^+\pi^0} \frac{2B(B^+ \rightarrow K^+\pi^0)\tau_{B^0}}{B(B^0
  \rightarrow K^+\pi^-)\tau_{B^+}} + A_{CP}^{K^0\pi^0}\frac{2B(B^0
  \rightarrow K^0\pi^0)}{B(B^0 \rightarrow K^+\pi^-)}.
\]
With data set from Belle II, the measurement precision will allow us
to compare the $A_{CP}$ predicted by this equation and the measured
$A_{CP}$ to check whether there is New Physics.




\subsection{$\tau$ Lepton Flavour Violation}

The Lepton Flavour Violation decays are highly depressed by SM, in a
branching fraction as $10^{-25}$, but they could be enhanced by some
New Physics scenarios. The red dots in Figure~\ref{fig:lfv} shows the
sensitivity for some LFV decays in Belle II~\cite{lfv}. The branching
fraction of the decays is within the capability of the experiment.


\subsection{Dark Sector}

\begin{figure}[htb]
\centering
\begin{minipage}{.45\textwidth}
\centering
\includegraphics[width=\textwidth]{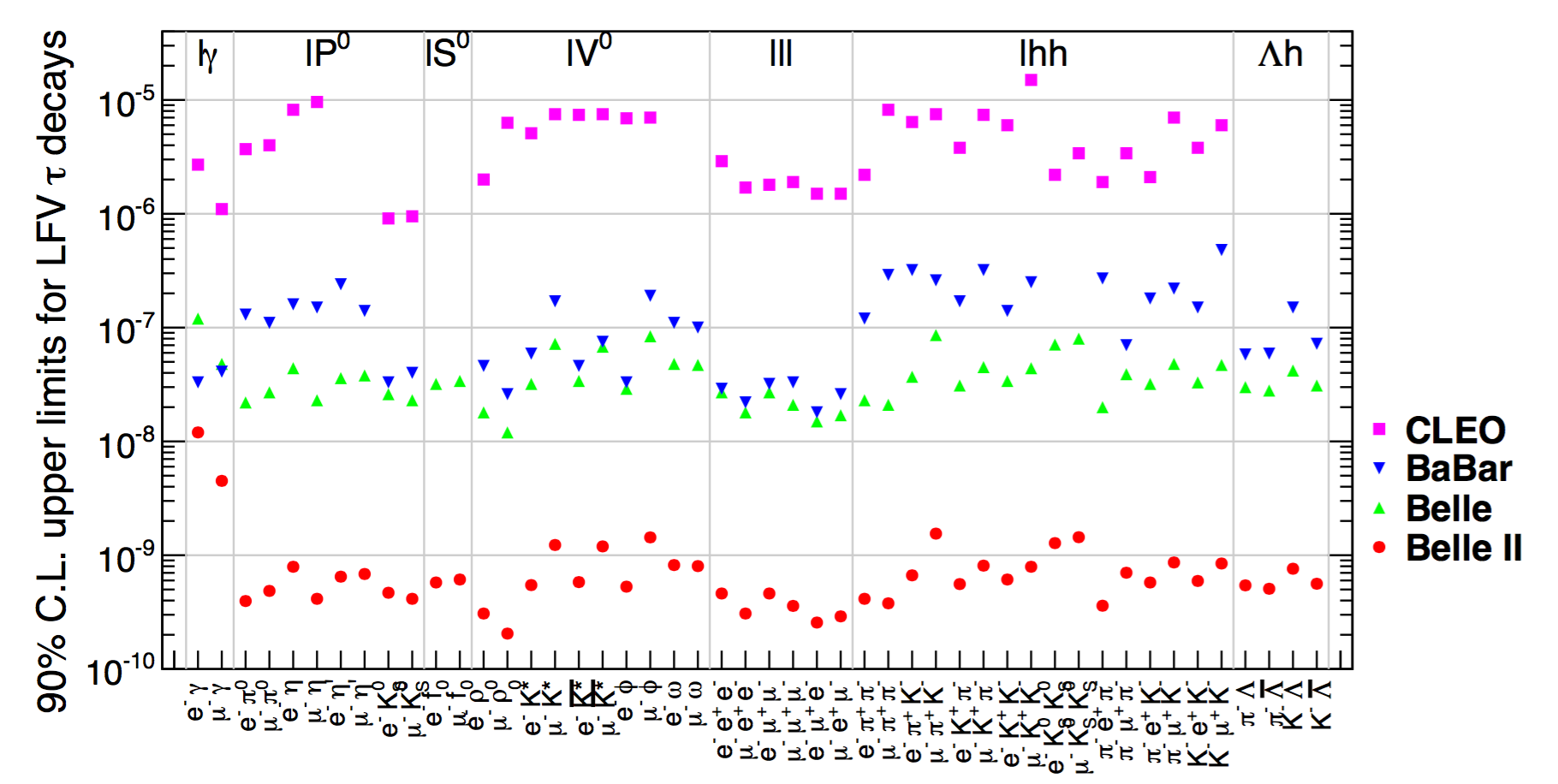}
\caption{The $\tau$ Lepton Flavour Violation measurement sensitivity
  by Belle II and other experiments.}
\label{fig:lfv}
\end{minipage}
\quad \quad
\begin{minipage}{.45\textwidth}
\centering
\includegraphics[width=\textwidth]{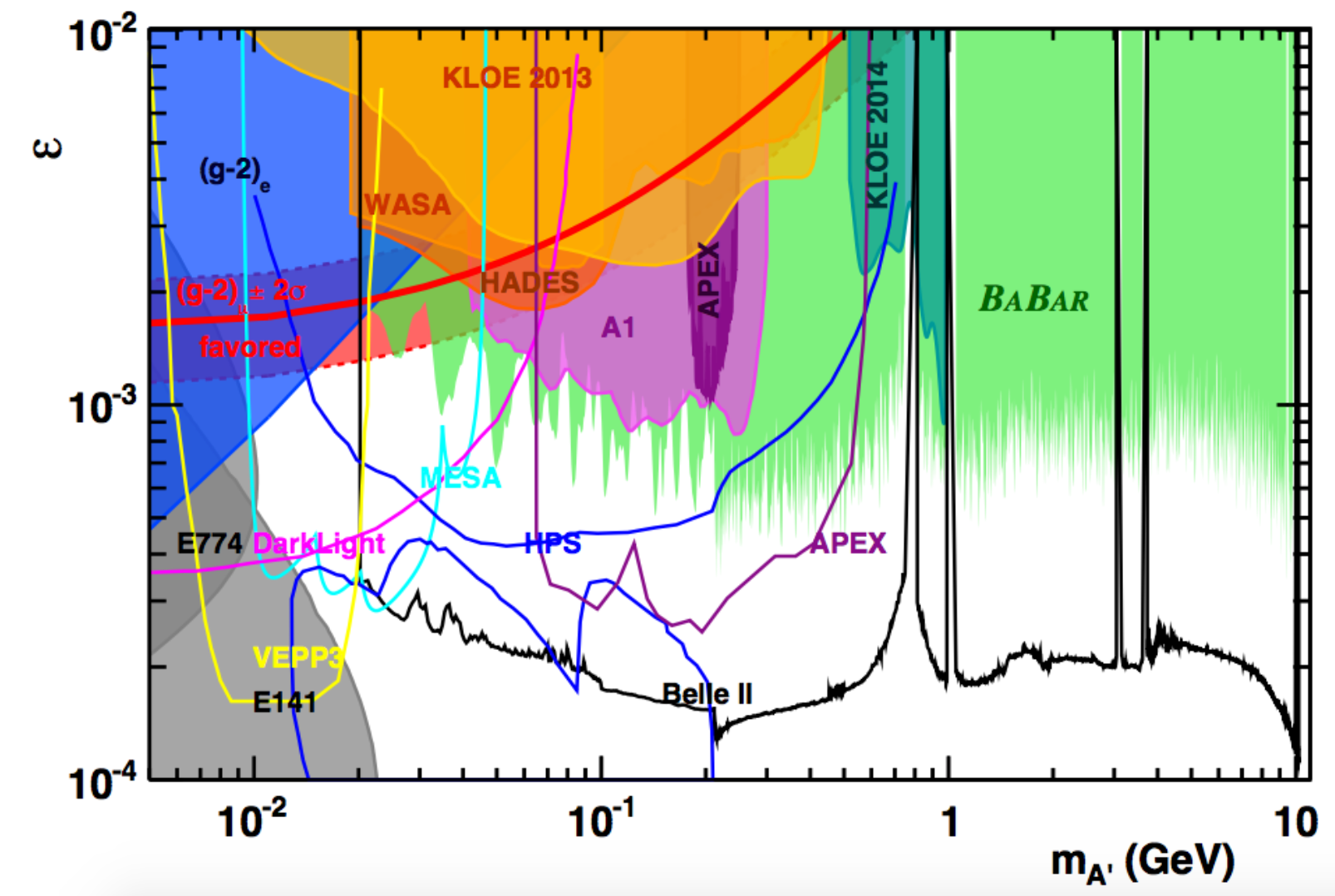}
\caption{The upper limit for dark photon measurement for different
  experiments.}
\label{fig:dark}
\end{minipage}
\end{figure}

It's possible to search dark matter in accelerator. One possibility is
the dark photon $A'$. Its mass is predicted in the range of MeV to
GeV. There're two ways to detect dark photon: probing leptonicaly
decaying dark photons through mixing, or probing sub-GeV dark matter
in invisible decays. The upper limits for dark photon measurement for
different experiments~\cite{dark} are shown in Figure~\ref{fig:dark}. With much
higher integrated luminosity, Belle II has an advantage to search dark
photon $A'$.

\section{Status and Schedule}

The SuperKEKB accelerator is now at the final stage of construction
and the upgrade of the Belle II detector is on going. There are three
phases in commission and operation of Belle II. In phase 1 at the
early 2016, the commissioning of various components will start without
rolling-in the detector. In phase 2 for middle 2017, Belle II detector
will be partly commissioned to take test physics data without the
vertex detector. Finally, in phase 3, the Belle II detector with full
apparatus is going to take physics data, which is expected at the end
of 2018. By 2024, the full data sample of 50 $ab^{-1}$ will be taken.

\section{Summary}

B factories like Belle and BaBar have proved to be excellent for
flavour physics. As an upgrade, the Belle II / SuperKEKB experiment
could play an important role in the searching for New Physics. With
the much larger data set, Belle II has a rich physics program ,which
makes it possible to study the channels with missing energy and
neutral particles in the final states. Now the accelerator and
detector are under construction, and the physics data taking will
start at the end of 2018.

\Acknowledgments

I'd like to thank the organizers of the DPF 2015 conference for
allowing me to give this talk. I'd also like to give my gratitude to
the Belle II collaboration and the HEP group of the University of
Cincinnati.

\end{document}